%% file: al2022.tex
\documentclass[format=acmsmall, nonacm, screen=True]{acmart}
\setcopyright{none}
\AtBeginDocument{%
  \providecommand\BibTeX{{%
    \normalfont B\kern-0.5em{\scshape i\kern-0.25em b}\kern-0.8em\TeX}}}

\usepackage{enumitem}
\usepackage{longtable}
\usepackage{adjustbox}

\usepackage{multirow}
\usepackage{array}
\newcommand{\PreserveBackslash}[1]{\let\temp=\\#1\let\\=\temp}
\newcolumntype{C}[1]{>{\centering\arraybackslash}m{#1}}
\newcolumntype{R}[1]{>{\raggedleft\arraybackslash}m{#1}}
\newcolumntype{L}[1]{>{\raggedright\arraybackslash}m{#1}}

\usepackage{multibib}
\newcites{nu}{Reference for supplementary literature}
\newcites{th}{References}

\setcitestyle{authoryear}

\usepackage{amsmath}
\usepackage{amsthm}

\usepackage{hyperref}
\hypersetup{hidelinks,
backref=true,
pagebackref=true,
hyperindex=true,
breaklinks=true,
colorlinks=true,
urlcolor=blue,
citecolor={blue},
bookmarks=true,
bookmarksopen=false,
pdftitle={Title},
}

\usepackage{xcolor}

\newif\ifshowcomment
\showcommenttrue
\ifshowcomment
\newcommand{\luyao}[1]{\textcolor{blue}{[luyao] #1}}
\else
\newcommand{\luyao}[1]{}
\fi

\begin{document}

\title{The Right Tool for the Job: Matching Active Learning Techniques to Learning Objectives}

\author{Sarah A. Jacobson}

\authornotemark[2]

\author{Luyao Zhang}
\orcid{0000-0002-1183-2254}

\authornote{Corresponding author: Luyao Zhang (email:lz183@duke.edu, institution: Data Science Research Center and Social Science Division, Duke Kunshan University, address: No.8 Duke Ave. Kunshan, Jiangsu 215316, China.)}
\authornote{Names by the alphabetical order of the last name. Sarah A. Jacobson: Williams College, Department of Economics, 24 Hopkins Hall Dr., Williamstown, MA 01267 USA. Luyao Zhang and Jiasheng Zhu: Duke Kunshan University, Kunshan, Jiangsu 215316, China}
\authornote{Also with SciEcon CIC, London, United Kingdom, WC2H 9JQ}

\author{Jiasheng Zhu}
\authornotemark[2]
\authornotemark[3]

\renewcommand{\shortauthors}{Sarah A. Jacobson and Luyao Zhang* and Jiasheng Zhu}

\begin{abstract}
Active learning comprises many varied techniques that engage students actively in the construction of their understanding. Because of this variation, different active learning techniques may be best suited to achieving different learning objectives. We study students’ perceptions of a set of active learning techniques (including a Python simulation and an interactive game) and some traditional techniques (like lecture). We find that students felt they engaged fairly actively with all of the techniques, though more with those with a heavy grade weight and some of the active learning techniques, and they reported enjoying the active learning techniques the most except for an assignment that required soliciting peer advice on a research idea. All of the techniques were rated as relatively effective for achieving each of six learning objectives, but to varying extents. The most traditional techniques like exams were rated highest for achieving an objective associated with lower order cognitive skills, remembering concepts. In contrast, some active learning techniques like class presentations and the Python simulation were rated highest for achieving objectives related to higher order cognitive skills, including learning to conduct research, though lectures also performed surprisingly well for these objectives. Other technique-objective matches are intuitive; for example, the debate is rated highly for understanding pros and cons of an issue, and small group discussion is rated highly for collaborative learning. Our results support the idea that different teaching techniques are best suited for different outcomes, which implies that a mix of techniques may be optimal in course design. (\textbf{JEL codes}: A20, A22, I21)

\noindent \textbf{keywords:} Active Learning Techniques, Student Engagement, Student Perceptions, Learning Objectives, Pandemic Pedagogy
\end{abstract}



\maketitle
\raggedbottom
\begin{acks}
We thank Jiaxin Wu and Tianyu Wu for their assistance in data collection. This research is approved by the Internal Review Board (IRB) at Duke Kunshan University. This research is jointly supported by the Assessment Grant from the Assessment Office and the Faculty Community for Active Learning Program, Center for Teaching and Learning, Duke Kunshan University.
\noindent We thank participants at the 2022 Pandemic Pedagogy Research Symposium and TeachECONference 2022 for helpful comments.
\end{acks}

\section{Introduction}

Active learning techniques have been shown to improve learning in classes at the university level. However, many instructors still make little or no use of active learning \cite{Sheridan2020}. Further, these techniques may vary in their effectiveness at achieving different goals \cite{Prince2004}, and while some past work studies different degrees of engagement in such techniques \cite{chi_2014_the}, more work is needed to explore the complex ways different techniques achieve desired outcomes. In this paper, we assess students’ perceptions of a variety of active learning techniques as implemented in an intermediate macroeconomics course, comparatively to each other and to a set of traditional teaching methods, with regard to a variety of course learning objectives including building skills for independent research.

Active learning techniques comprise a wide variety of alterations to traditional modes of instruction, ranging from small exercises like think-pair-share \cite{kaddoura2013think} or “Do Now” \cite{collins2020now} to full transformations like flipped classrooms (e.g.,~\cite{caviglia2016flipping}). The most common definition of active learning is instruction that “engages students in the process of learning through activities and/or discussion in class, as opposed to passively listening to an expert” \cite{freeman2014active}. A seminal meta-analysis of experimental studies \cite{freeman2014active} found robustly that across STEM disciplines, student performance and retention were significantly improved by active learning methods. Some of this evidence points to greater benefits from active learning with regard to higher order cognitive skills rather than lower-level skills like memorization \cite{Omelicheva2008}. 

The rest of the paper is organized as follows. In the next section, we describe the teaching techniques we study. In the following section, we describe the survey and the context in which it was deployed (the survey itself is in Appendix A), and present some hypotheses grounded in the results of past studies. Next, we present summary statistics of the survey data, and next, results from analyzing the data. Finally, we conclude.

\section{Description of Techniques}

Table~\ref{tab:1} presents the ten teaching techniques we study in this paper, along with whether and to what extent the technique was graded in the course in which we studied them.\footnote[1]{When each lesson or assignment was introduced, the pedagogical rationale for how the teaching technique would benefit a particular learning objective was not explicitly explained.} Below, we describe each activity in detail. We group them into three categories: “traditional” techniques, which are mainstays of traditional classrooms; “active” techniques, which are active learning elements that have been discussed in STEM education for decades, as the citations below show; and “custom” techniques, which are implementations of novel exercises that were developed specifically for this class.

\input{tables/Table1}

\newpage

\noindent \textbf{Traditional:}

\begin{enumerate}[label=\textbf{\alph*}.]
    \item  \textbf{Lecture:} Students are mostly passive while the instructor lectures. In our implementation, the lectures and students were in-person, and lecturing accounted for 60\% of class time. 
\item \textbf{Problem sets:} Traditional graded take-home assignments. In our implementation, they included short essay questions and interaction with the Python simulations (technique i). 
\item \textbf{Quizzes / exams:} Traditional graded individual assessments; in our implementation, quizzes happened weekly online and exams happened during the exam week in class, and all were multiple-choice.

\end{enumerate}

\noindent \textbf{Active:}

\begin{enumerate}[label=\textbf{\alph*}.]
\setcounter{enumi}{3}
    \item  \textbf{Written reflection:} This has two components. First, students make weekly comments in the learning management system about the subject matter, what they learned, and questions they have. Second, in one problem set, students reflect on the classroom game (technique j) by describing a research question the game could be used to study and considering how they and others might behave in a study of that research question. In a reflective assignment like this, students have the opportunity to construct and explore the boundaries of their understanding of concepts \cite{knoblauch_1983_writing}. Reflection can take many forms, including individual written reflection assignments like those in this class \cite{chan2021students}. Reflection of this type has been used in economics for decades to introduce active learning \cite{crowe_1986_using}, to personalize course content \cite{brewer_2006_making}, and to give instructors a window into student thought processes \cite{goebel_2019_recounting}.
\item  \textbf{Research idea presentation:} Each student gives a brief presentation to the class on their own research idea inspired by a common prompt. Presentations are used often in economics classes in which students perform independent work; most often, as in \cite{morreale_2020_creating}, it is a presentation of a completed work rather than an early-stage idea as in our case.
\item  \textbf{Small group discussions:} In class, students are randomly matched into groups of 4 to 6 students and are asked to discuss a prompt asking them to apply macroeconomic theory to a specific real-world issue. Each group then shares their takeaways with the whole class. Small group work is often used in collaborative learning projects (e.g.,\cite{johnston_2000_an, Yamarik2007}) and flipped or partially flipped classrooms (e.g., \cite{roach_2014_student}). However, students may be resistant to the use of class time for such discussions unless convinced of its utility \cite{clinton2017student}.
\item  \textbf{Debate:} Students debate the pros and cons of different approaches to macroeconomic policy (e.g., rules vs. discretion) or different schools of thought (e.g., neoclassical vs. new Keynesian). Debates have been used to teach for millennia \cite{kennedy2009power} and are a low-tech way to engage students to many ends, including to promote critical thinking and reinforce learning  \cite{vo_2006_debate}, explore normative and ethical questions \cite{hennessey_2014_motivating}, and air pluralist approaches \ \cite{obengodoom_2019_pedagogical}, and have been found effective in helping students learn (e.g., \cite{kennedy2009power}).
\item  \textbf{Peer advice:} Students seek advice from other students on the research idea they will present to the class (see technique e). In our implementation, it is up to the student how and in what form to get that advice, but students must incorporate advice and responses to it in the presentation. Peer feedback, especially on written work (e.g., \cite{cohen_2019_scalable}), is a common active learning technique, and has been shown repeatedly (e.g., \cite{cho_2010_peer}) to benefit both the giver and receiver of feedback. However, most studies focus on summative feedback rather than formative feedback like this \cite{mulder2014peer}.

\end{enumerate}

\newpage

\noindent \textbf{Custom:}

\begin{enumerate}[label=\textbf{\alph*}.]
\setcounter{enumi}{8}
    \item \textbf{Python simulations:} The instructor uses macroeconomic models programmed in Python Jupyter Notebook to run simulations, demonstrating the models in class, and students interact with the model by changing parameters in take-home assignments. Simulations can help students understand how models work and the different outcomes that can be expected in different scenarios, and economists have used them in the classroom to help students learn about growth and inequality \cite{hanlon_2013_integrating}, risk and time preferences and how finance and insurance markets work  \cite{guo_2014_demystifying}, and imperfect competition \cite{gorry_2015_numerical}.
\item \textbf{Game:} Students play a cryptocurrency investment game (described in detail in \cite{zhu_2022_game} that explores the tradeoffs between rule-based and discretionary policies in the context of cryptocurrency and deviations from neoclassical economic theory with regard to investment behavior. Games have been a mainstay of active learning in economics classrooms for decades; \citeN{vogt-1999} championed their use and recent examples include \citeN{abidoye2021seeds}.
\end{enumerate}

\section{Survey and Hypotheses}

We incorporated the ten teaching techniques described in Section II into an intermediate macroeconomic course at Duke Kunshan University in Spring 2021. At the end of the last lecture, we delivered an online survey to students (see Appendix A for the survey) to assess their perceptions of how engaging each technique was, how much they enjoyed each, and how effective each was at contributing to a set of learning objectives. The survey was implemented in Qualtrics. Students knew their responses would be anonymized by the University Center for Teaching and Learning, which helped conduct the survey, before being shared with the researchers. Students received 40 RMB (about \$6.30 USD) for completing the survey and could elect to prevent their responses from being used in this study. Of the 32 students in the class, we received 28 complete survey responses.

The survey first asks, on a 7-point Likert scale, how actively the student engaged with each of the ten techniques. It then asks, on a 7-point Likert scale, how effective the student thought each technique was at achieving each of six learning objectives: remembering concepts, understanding different perspectives on macroeconomic issues, analyzing real-world issues, collaborative learning, conducting systematic research, and understanding the pros and cons of macroeconomic policy. Students then rank the techniques by how much they enjoyed them and separately by how useful they felt each was in developing their Signature Work (the name in this academic program for independent research projects all students must complete in their senior year). Finally, in a free text response question, the survey asks students to give an example of how the learning techniques were helpful.

We develop several hypotheses based on the results of past studies. First, we hypothesize the techniques that are more active (written reflection, presentation, peer advice, small group discussion, debate, Python simulation, and game) to be rated as more engaging than those that are less active (lecture, quizzes / exams, problem sets). However, this pattern may be altered by the students’ incentives: higher grade weights may also make them engage more actively with a technique. We also expect that students will report enjoying the active techniques more (even if they carry less grade weight).

We might expect that techniques that are more active also better achieve learning outcomes, in general. Since our outcomes of interest are self-reported subjective measures of learning, however, students’ reported effectiveness may be altered by the tendency noted in \cite{deslauriers_2019_measuring} for students to feel like they learn more in passive learning modes even though they actually learn more in active learning modes.

Next, we hypothesize that different techniques will be effective at achieving different learning outcomes. Techniques that give the instructor more control, notably lecture, quizzes and exams, and problem sets, may be the best at helping students remember specific concepts. Those that allow many voices to be heard, such as presentations, peer advice, small group discussion, and debate, should be most effective at helping students understand different perspectives. Those that are structured around concrete macroeconomic policy problems from the real world, such as the presentation, peer advice (which relates to the presentation), small group discussion, and debate should be experienced by students as most effective in helping them learn to apply macroeconomics to understand real-world issues and the pros and cons of macroeconomic policy. Techniques in which students work together, such as peer advice, small group discussion, and debate, should be most effective for collaborative learning. Learning to conduct research in macroeconomics should be most affected by techniques in which students are encouraged to think about research, namely the presentation, peer advice, and the game.

\section{Data}

In this section, we report summary statistics from the survey. In the following section, we will perform statistical tests to compare students’ perceptions of the techniques.

\begin{figure}[H]
    \centering
    \includegraphics[width=0.9\textwidth]{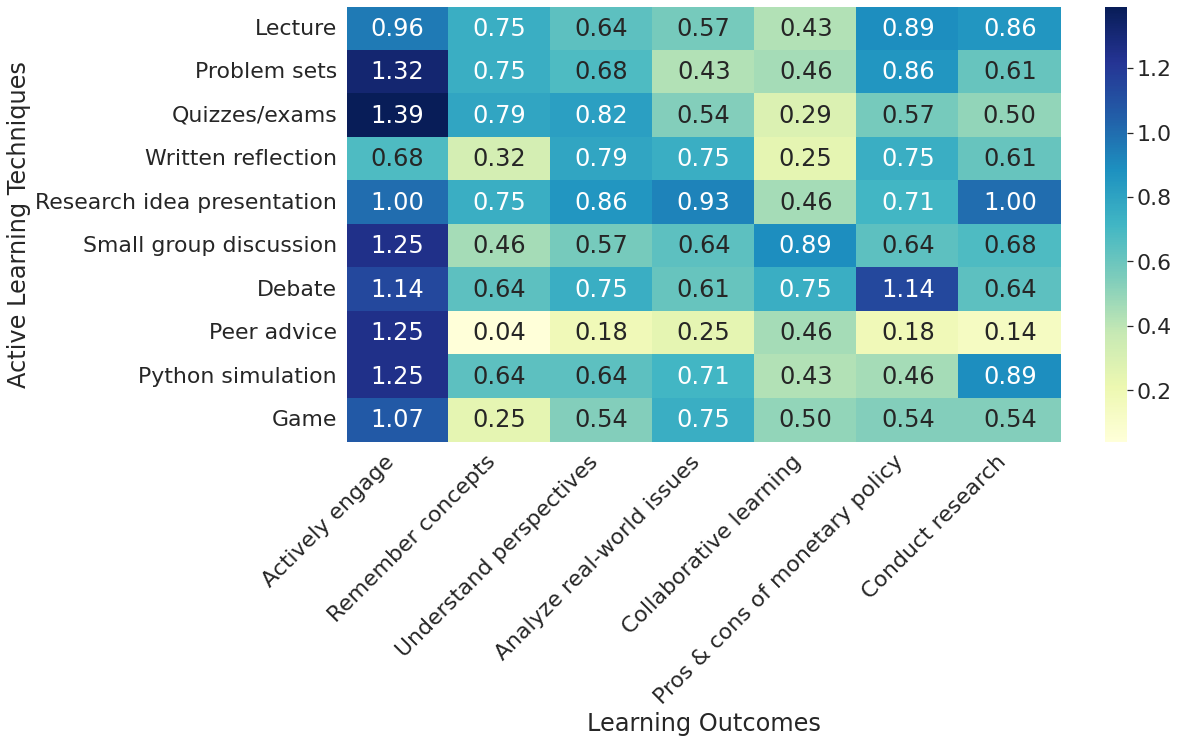}
    \caption{Reported Engagement and Learning Outcome Effectiveness}
    \label{fig:1}
\end{figure}

Figure~\ref{fig:1} reports the extent to which students felt they engaged actively with each technique, and to which they felt each technique was effective with regard to the learning outcomes we queried them about. Positive values indicate a student felt the technique was engaging or effective, while negative values indicate they felt they didn’t engage with it very actively or it was ineffective.

Note: Numbers in (and darkness of) cells indicate average responses to the question; the seven options for each question are coded from -3 to +3. For row 1, -3 means that they engaged “Extremely inactively” with the technique, 0 means “Neither actively or inactively,” and +3 means “Extremely actively;” for the rest of the table, -3 corresponds to “Extremely ineffective” and +3 corresponds to “Extremely effective” at achieving the named learning objective. N=28.

Figure~\ref{fig:1} yields a few lessons. First, students considered all of the techniques relatively engaging and effective at all of the outcomes; there is no value reported less than zero. Second, there is no one technique that seems to be the “best” or the “worst:” each has some outcomes for which more and fewer students find it to be effective. We will explore the results in detail and test this variation in the next section.

\section{Results}
We now test the hypotheses we proposed in Section III. We first discuss how actively students said they engaged with, and how much they enjoyed, the techniques. We then discuss how effective the students reported the techniques were at achieving the learning outcomes, separating out and saving for last our discussion of how students felt the techniques helped them learn how to conduct research.

\subsection{Engagement and Enjoyment}

First, Figure~\ref{fig:1} shows that students reported engaging most actively with quizzes and exams, and next with the game, peer advice, and the debate, and the least with the written reflection. Table~\ref{tab:2} shows that some of these differences are significant: the traditional tools of problem sets and quizzes / exams and the active technique of small group discussion are equally rated as the most engaging, and the active learning technique of written reflections is rated as least engaging. We had hypothesized that the active learning techniques would be more actively engaged with, but this is not entirely borne out in the data. This appears in part to be because, as we had conjectured, the hefty grade weights of quizzes, exams, and problem sets forced a form of engagement, though it also appears that students simply find written reflections unengaging.

\input{tables/Table2}

\begin{figure}[H]
    \centering
    \includegraphics[width=0.8\textwidth]{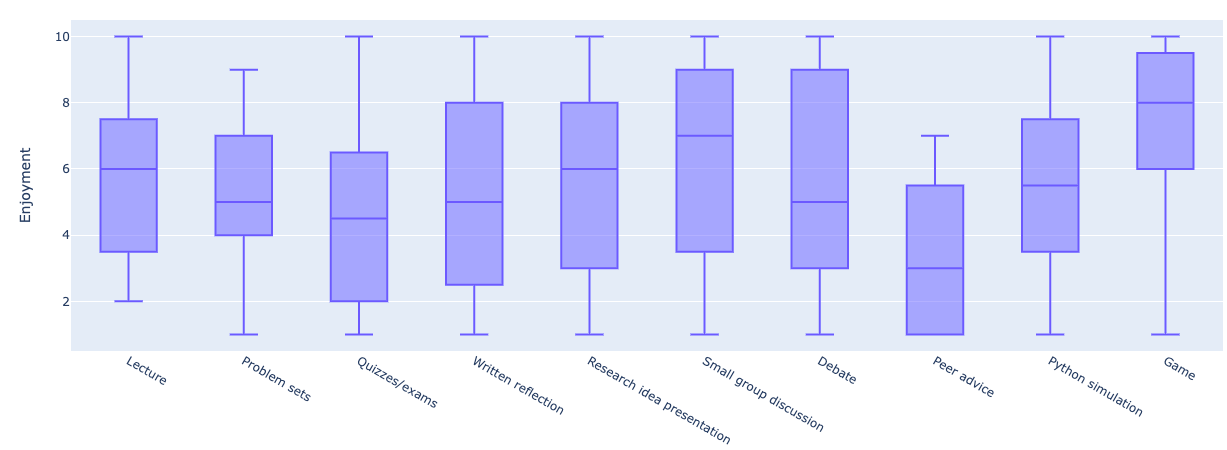}
    \caption{Enjoyment of the Teaching Techniques}
    \label{fig:2}
\end{figure}

Next, Figure~\ref{fig:2} shows that students enjoy the game the most and peer advice the least. Table~\ref{tab:3} shows that these differences are significant, and also shows some daylight between other techniques: namely, quizzes and exams are less enjoyable than lecture or small group discussion. The relatively poor performance of peer advice is not surprising, given that other studies have found that students have mixed perceptions of peer review assignments \cite{mulder2014peer}. Thus, students enjoyed many of the active learning techniques the most, but not all of them; notably, students engaged very actively with the game and enjoyed it the most of all the techniques, whereas they engaged very actively with the peer advice and enjoyed it the least of all the techniques.

Note: Students ranked each of the 10 activities in order of how much they enjoyed them. This figure reports ten minus the rank score (e.g., the item ranked 2nd most enjoyable would show a high value of 8 in this graph, while that ranked 9th most enjoyable would be a low value of 1). For each activity, the horizontal line indicates the median of student rankings, the box represents the interquartile range, and the whiskers represent the range of responses.

\input{tables/Table3}

Instructors may find active engagement and enjoyment important ends in and of themselves. In addition, engagement and enjoyment may be important for learning. While we cannot claim causality, it is suggestive that how actively a student said they engaged with a technique and how effective they felt it was at achieving each learning objective were strongly correlated (Spearman p < 0.03 in each case except p = 0.07 for the presentation achieving the objective of understanding of the pros and cons of macroeconomic policies, see Table B1 in Appendix B). Surprisingly, enjoyment is not broadly correlated with learning objectives (see Table B2 in Appendix B): it is for peer advice for several learning objectives, which could relate to either heterogeneous attitudes or heterogeneous experiences generating both enjoyment and a positive learning experience from the activity; it also is for quizzes and exams for several objectives, and the game for understanding the pros and cons of macroeconomic policies. In other words, the activities that students gave the lowest average enjoyment rankings to were also the ones that showed some correlations between enjoyment and learning objectives. We also note, perhaps even more surprisingly, that enjoyment and engagement are almost uncorrelated (see Table B3 in Appendix B); there is only a significant correlation for the research idea presentation.

These results show that the activities that students reported feeling more actively engaged in and, to a much lesser extent, the ones they reported enjoying the most were also the ones they viewed as most effective in achieving learning objectives.

\subsection{Matching Active Learning Techniques to Learning Objectives}

We next, for each learning objective, make pairwise comparisons of students’ reported effectiveness of each technique using Wilcoxon rank sum tests. Results are reported in Table~\ref{tab:4}. We find that some, but not all, of the differences in students’ reports of the effectiveness of individual techniques for achieving learning objectives observed from Figure~\ref{fig:1} are statistically significant. 

\input{tables/Table4}

First, the learning objective associated with the lowest order cognitive skills, remembering concepts, is seen by the students as well served by traditional teaching techniques: lecture, problem sets, and quizzes and exams are rated as the most effective for this outcome. However, the Python simulations, research idea presentation, and debate also performed relatively well for this objective. Students rate peer advice as the least effective for memorization.

Students report feeling that a variety of techniques across the three categories (traditional, active, and custom) are effective at helping them understand different perspectives in macroeconomics. We conjecture this is because this outcome requires learning basic facts (a lower level cognitive skill) as well as making connections about how and why perspectives emerged and making judgments about them, which requires higher order cognition. The only technique that is reported as significantly less effective for this objective is seeking peer advice regarding a research idea presentation, which is not surprising given that students were seeking peer advice about a research idea, not about comparative approaches to macroeconomics.

For analyzing real-world issues, the technique that dominates the most other techniques is the research idea presentation, which is intuitive because students were instructed that the research idea should relate to real-world issues. Some of the rest of the active and custom tools (the written reflection, Python simulations, and game) were also reported to be very effective on this dimension. Peer advice was not rated as very effective for this outcome, and problem sets also performed relatively poorly. The poor performance of peer advice on this objective is somewhat surprising given that the peer advice was about the research idea presentation.

The techniques students reported as most effective for collaborative learning were, unsurprisingly, active learning activities in which they worked with other students: small group discussion and debate dominate most of the alternatives. The written reflection, a solitary activity, is unsurprisingly the least collaborative. Of the other techniques, only the game was ranked as being better for collaborative learning than any other technique. It is notable that while the peer advice assignment explicitly requires students to work with others, students do not rate it as one of the most effective techniques for collaborative learning, though it doesn’t perform as badly with regard to this objective as it does with regard to the other objectives.

For understanding the pros and cons of macroeconomic policy, student responses show the debate dominates the other techniques, which is intuitive since a debate specifically airs pros and cons. Other techniques the students report are particularly effective for this objective are lectures and problem sets, though the latter is rated as less effective than the debate. The peer advice assignment is rated as less effective than most of the other techniques, which may again be because students were giving advice about research rather than policy. The Python simulations also were not rated as highly as other techniques, perhaps because these are more about models (specifically, how starting conditions affect economic growth) than policy.  

All things considered, our results can be summarized as follows. First, while all of the techniques are considered relatively effective for all of the outcomes, the written reflection and especially the peer advice assignment are most commonly ranked as least effective techniques. Second, while the lecture, problem sets, and quizzes / exams are reported to be quite effective for many outcomes (which is congruent with past research showing that students feeling of learning may be greater for passive methods of instruction; see \citeN{deslauriers_2019_measuring}), and are the best for memorization of concepts, they are not as effective as specific other more innovative techniques for outcomes to which those techniques have a clear link: the research idea presentation for analyzing real-world issues, the discussion and debate for collaborative learning, and the debate for understanding pros and cons of macroeconomic policy.

\subsection{Effectiveness of Techniques for Learning to Do Research}

\begin{figure}[H]
    \centering
    \includegraphics[width=0.8\textwidth]{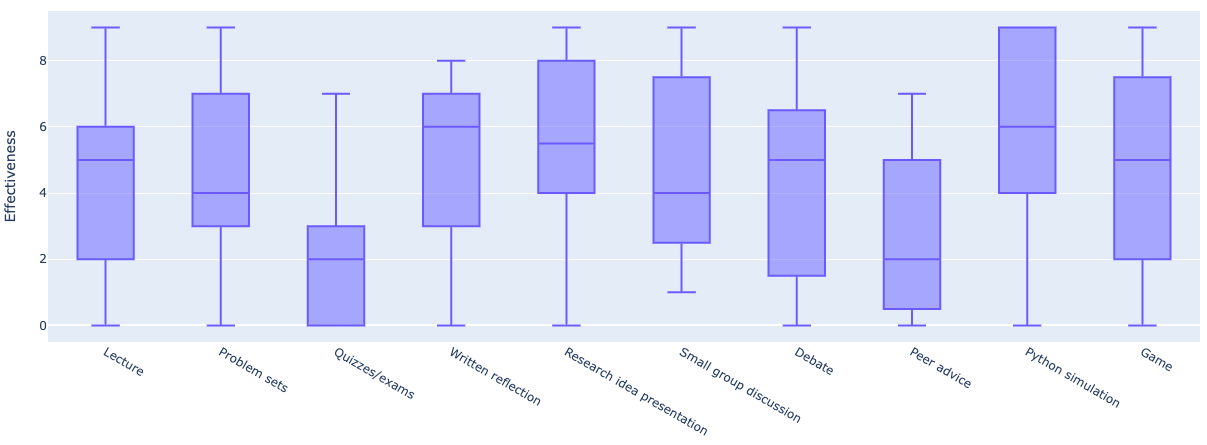}
    \caption{Rankings of Activities’ Effectiveness in Developing Research Project}
    \label{fig:3}
\end{figure}

We have two sources of student responses regarding learning to do research: we asked students how effective each technique was at the learning objective of learning how to do research, results from which we report in Table~\ref{tab:5}, and we asked students to rank the activities with regard to how useful they were in conducting their own research projects, rankings we show in Figure~\ref{fig:3} and Table~\ref{tab:6}.

\input{tables/Table5}

In Table~\ref{tab:5}, we show results of tests comparing students’ self-reports regarding the effectiveness of the different teaching techniques for learning to conduct systematic research. Lecture and the research idea presentation are both rated as relatively effective for this outcome. Since the research idea presentation is a step toward conducting research, it is not surprising that this assignment supports this outcome. The Python simulations are rated as significantly more effective than some other techniques as well, perhaps because some students hoped to use Python in their research; indeed, as we show in a word cloud in Figure B1, in free text responses to a question about how the learning activities helped them, the word “Python” is one of the most often-used words (second only to “research”). The poor reported performance of the peer advice assignment is again notable, since the advice is about the research idea presentation. It appears that students feel more comfortable gathering ideas relevant to their research from the instructor in lecture than from peers in an informal feedback setting.

Note: Students ranked each of the 10 activities in order of how useful they were in helping them develop their Signature Work, which is the name in this academic program for a student’s culminating research project required for the degree. We report here ten minus the rank score (e.g., the item ranked 2nd most useful would be a high value of 8 in this graph, while that ranked 9th most useful would be a low value of 1). For each activity, the horizontal line indicates the median of student rankings, the box represents the interquartile range, and the whiskers represent the range of responses.

Results differ only slightly when we look at students’ rankings of which activities were most useful to them in developing their Signature Work (the required culminating research project for the degree). As Figure~\ref{fig:3} demonstrates visually and Table~\ref{tab:6} shows statistically, congruent with the results above, the Python simulations are ranked consistently higher than many other techniques, and the research idea presentation is ranked more helpful than lecture, while the peer advice and quizzes and exams are ranked lowest. This is again congruent with the word cloud shown in Appendix Figure B1. The most notable difference between the effectiveness scores in Table~\ref{tab:5} and the rankings in Figure~\ref{fig:3} and Table~\ref{tab:6} is that lecture performs quite well in in rated effectiveness at the learning objectiveness of learning to conduct systematic research, but does not dominate many alternatives in the ranking of usefulness for preparing the student’s research project.

\input{tables/Table6}

\section{ Conclusion}

Active learning has been studied for decades and found to often be more effective than traditional expositional teaching methods, especially for achieving higher order learning outcomes. However, active learning techniques vary tremendously in their focus and mechanics, and therefore they can be expected to vary in what learning objectives they are most effective at achieving. We explore this variation by deploying a set of ten teaching techniques, some traditional and some active, and surveying students about how actively they engaged with the techniques, how much they enjoyed them, and how effective each technique was at achieving each of six learning objectives, with a particular interest in teaching students to conduct research. We hypothesized that the traditional techniques would be most effective at achieving lower level outcomes like remembering concepts while active techniques would be most effective at achieving objectives related to higher order cognitive skills like learning to do research, though we recognized that assignments that have a large impact on student grades could “engage” students in the sense of getting their attention and making them work hard in service of a grade. We also conjectured that some techniques were simply better matched to particular outcomes.

The responses of students in our sample supported most of these conjectures and yielded a number of other intuitive matches between teaching techniques and learning objectives, but also returned some surprising results. For example, students reported lectures as particularly helpful for most of the outcomes, including learning to do research, though this might reflect students’ preference for passive learning because of the additional cognitive effort that active learning requires \cite{deslauriers_2019_measuring}. As another example, students rated an assignment in which they had to seek advice from a peer relatively low on most objectives.

Our results are limited by our sample size (only 28 students from one semester of one course) and our use of only student self-reported responses rather than more objective measures like grades, though we argue that these self-reports let us get at some nuanced concepts that objective measures like test scores would fail to capture. Further, it seems very likely that exactly how any technique is implemented would significantly alter what outcomes it achieves and how effective it is at achieving those outcomes, so we can only claim to report on the perceived effectiveness of our implementation of these techniques. For example, if the assignment in which students obtained advice from peers about a research idea presentation was implemented differently, perhaps it would perform better. 

However, many of our results are intuitive and therefore likely generalizable. Overall, we hope that our results highlight the heterogeneity of learning techniques’ effects to help instructors fine tune their adoption of new techniques to serve specific purposes. 

\newpage


\bibliographystyle{ACM-Reference-Format}
\bibliography{al2022.bib}

\nocite{*}

\newpage
\appendix

\section{ Survey}

{\textbf{Assessment Survey for Students’ Engagement and Perception}}

\subsection{Consent Form}

\begin{enumerate}[label=(\alph*)]
    \item Key Information

We are learning too!

You are being asked to participate in a research study being conducted by Dr. Luyao Zhang, Assistant Professor of Economics and Senior Research Scientist at Data Science Research Center at Duke Kunshan University. The purpose of this study is to understand and improve the process of classroom teaching and the learning experience. 
We are asking you to share your thoughts in a reflection survey.

\item	Voluntariness and Confidentiality

Your participation is completely voluntary, and you may withdraw at any time. Your decision will have no impact on your grades in this or any other course you have taken or will take in the future. Your professor will not know whether or not you agreed to share your information. DKU’s CTL office will hold all the records of which students consented to participate in this project and will only provide Dr. Luyao Zhang data from students who consent. 

If you choose not to share your survey data, identifying information will be used to remove your data. Identifying information will then be removed by the DKU CTL office prior to Dr. Luyao Zhang analyzing the data. DKU CTL will store the original data in Duke box and it will be deleted after one year. De-identified data may be used for research in education studies and be included in research publications.

\item	Release of Course Data for Research

We are asking that you release to us your data from the following reflection survey to be used in research. Your data will be used to help us better understanding the teaching and learning experiences here at DKU. We are absolutely not making any judgments about any individual participants. None of your personal information will be included in any analysis or publications of our findings.

\item	Contact Information

For questions about this research, please contact Luyao Zhang at lz183@duke.edu.

If you agree to be in this study but later change your mind and want to withdraw, please contact Center for Teaching and Learning at dku\_ctl@dukekunshan.edu.cn. 

For questions about your rights as a participant in this research, please contact the Duke Kunshan University IRB at dku-irb@dukekunshan.edu.cn. 
\end{enumerate}

Please select one of the following options:

\begin{itemize}
    \item	Yes, I consent to share my data from this course for research purposes
\item	No, I do NOT want to share my data from this course for research purposes
\end{itemize}

The next screen may ask you to enter your NetID to confirm your selections. If you are already logged in, you may not see a Duke login screen, but your responses will be recorded. Please print this page before proceeding if you would like a copy for your records.

\subsection{Survey Questions}
Please answer the following question based on your experience in ECON 204 Intermediate Macroeconomics Spring 2021.

\begin{figure}[H]
    \centering
    \includegraphics[width=\textwidth]{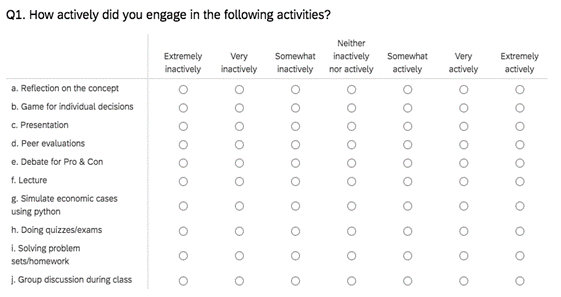}
\end{figure}

\begin{figure}[H]
    \centering
    \includegraphics[width=\textwidth]{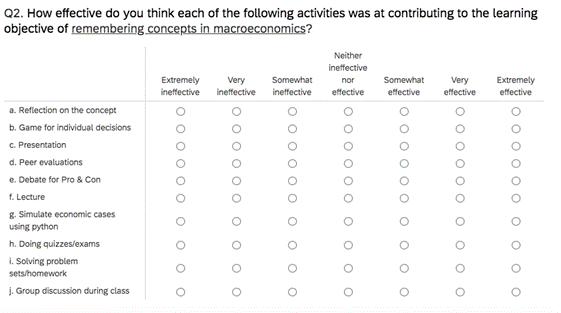}
\end{figure}

\begin{figure}[H]
    \centering
    \includegraphics[width=\textwidth]{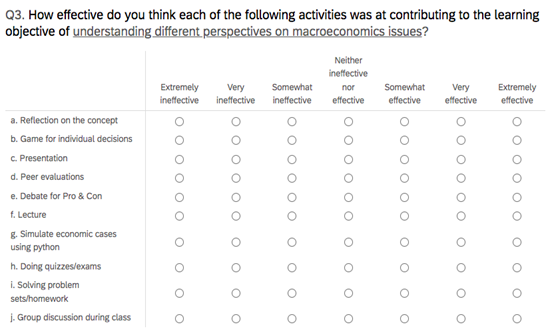}
\end{figure}

\begin{figure}[H]
    \centering
    \includegraphics[width=\textwidth]{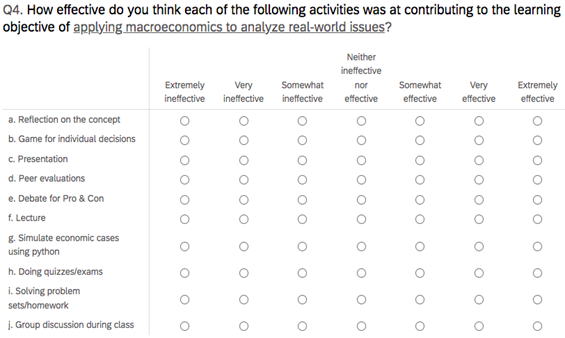}
\end{figure}

\begin{figure}[H]
    \centering
    \includegraphics[width=\textwidth]{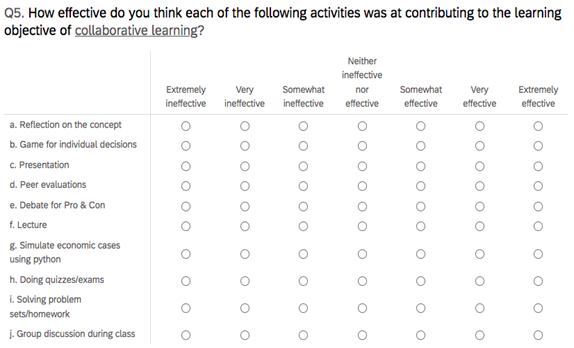}
\end{figure}

\begin{figure}[H]
    \centering
    \includegraphics[width=\textwidth]{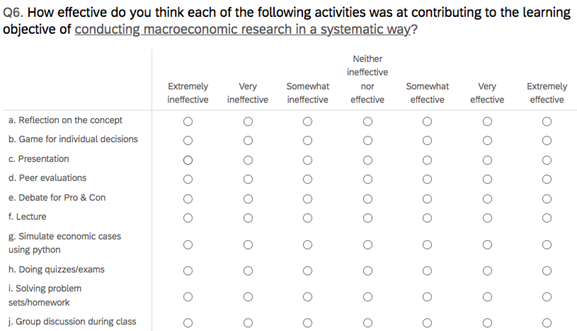}
\end{figure}

\begin{figure}[H]
    \centering
    \includegraphics[width=\textwidth]{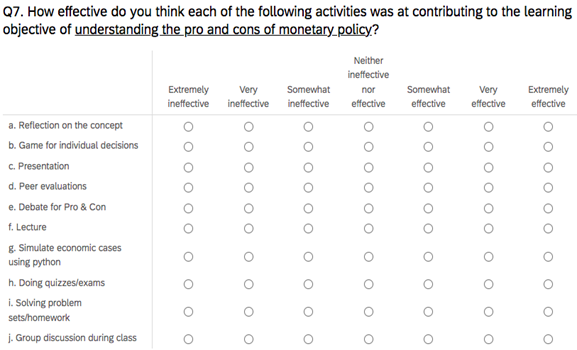}
\end{figure}

\begin{figure}[H]
    \centering
    \includegraphics[width=\textwidth]{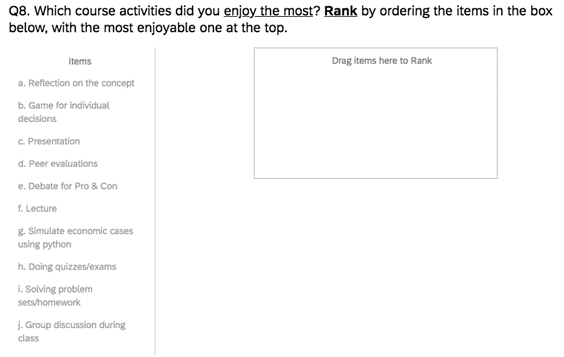}
\end{figure}

\begin{figure}[H]
    \centering
    \includegraphics[width=\textwidth]{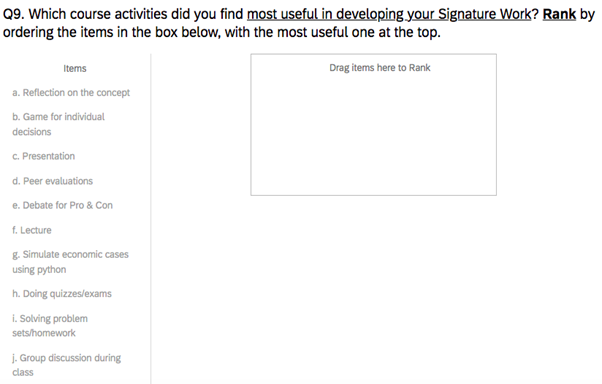}
\end{figure}

\begin{figure}[H]
    \centering
    \includegraphics[width=\textwidth]{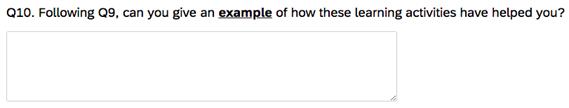}
\end{figure}

\section{Supplementary Figures and Tables}

\input{tables/Table7}

\input{tables/Table8}

\input{tables/Table9}

\input{tables/Table10}

\input{tables/Table11}

\begin{figure}[H]
    \centering
    \caption{Word Cloud of Students Responses to the Question “Can you give an example of how these learning activities have helped you?” (dark version)}
    \includegraphics[scale=.17]{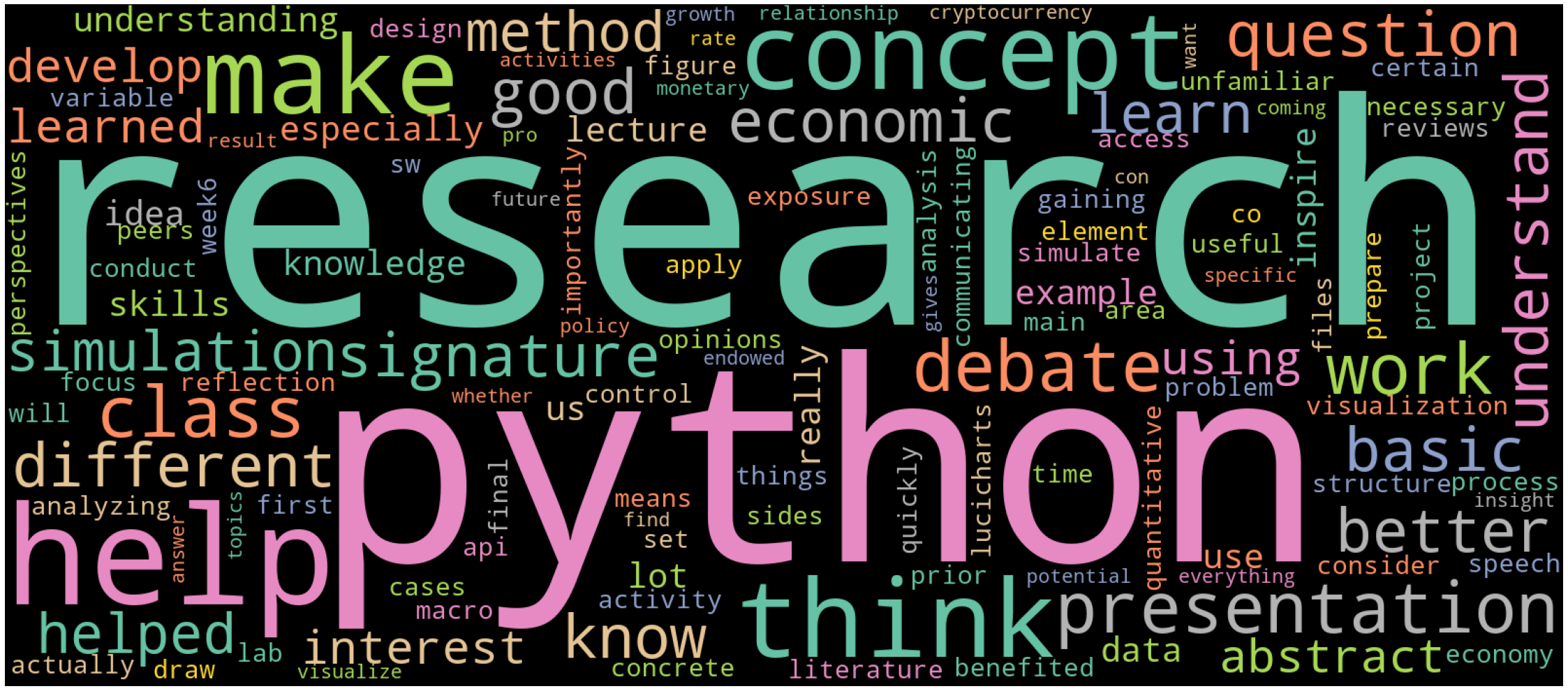}
\end{figure}

\begin{figure}[H]
    \centering
    \caption{Word Cloud of Students Responses to the Question “Can you give an example of how these learning activities have helped you?”(light version)}
    \includegraphics[scale=.17]{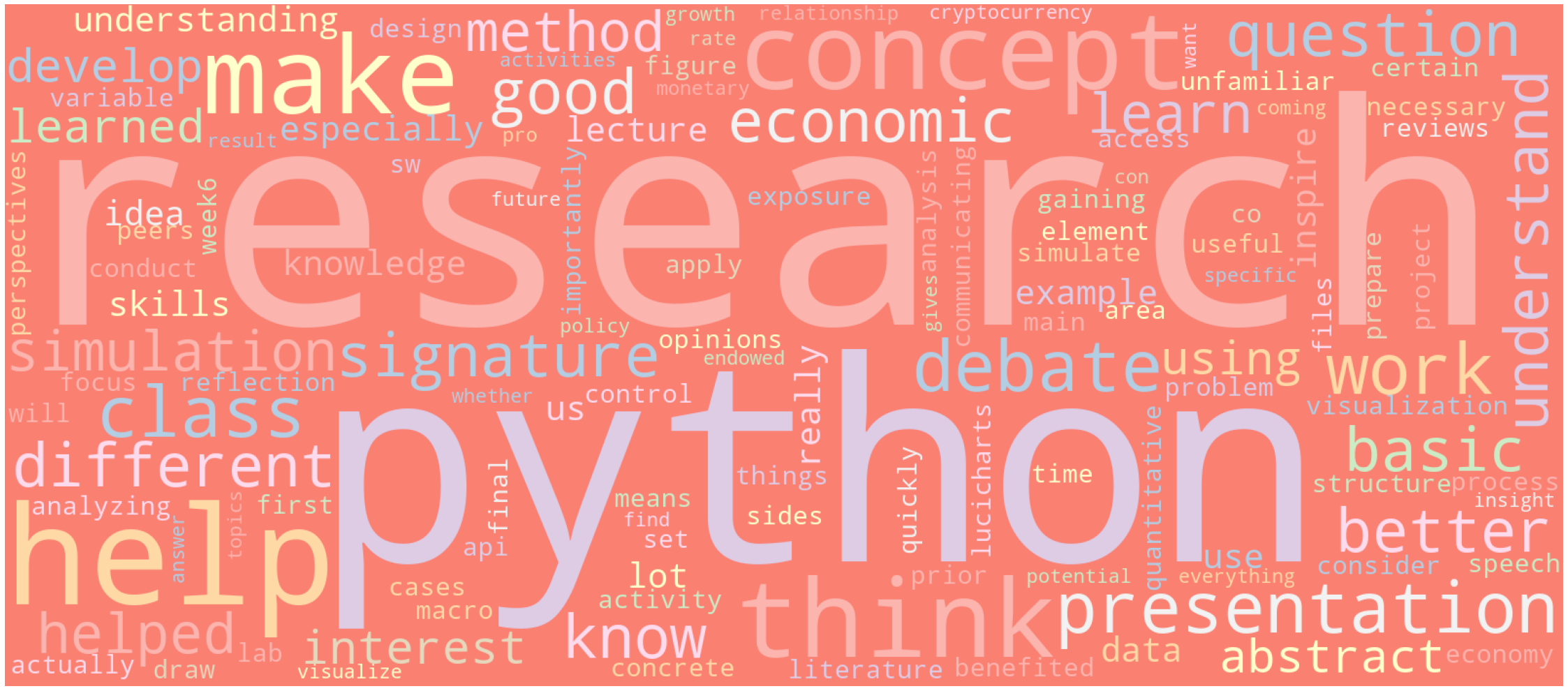}
\end{figure}

\end{document}

%% file: tables/Table1.tex
\begin{table}[H]
\centering
\caption{Techniques Studied}
\label{tab:1}
\resizebox{0.85\textwidth}{!}{%
\begin{tabular}{llcc}
\hline
 &
   &
  \begin{tabular}[c]{@{}c@{}}Grade Weight if \\ Directly Graded\end{tabular} &
  \begin{tabular}[c]{@{}c@{}}Grade Weight of Assignments\\  Referring to Exercise\end{tabular} \\ \hline
\multirow{5}{*}{Traditional} &
  a. Lecture &
  - &
  - \\
 &
  b. Problem   sets &
  40\% &
  - \\
 &
  c. Quizzes /   exams &
  40\% &
  - \\
 &
  d. Written  reflection &
  \begin{tabular}[c]{@{}c@{}}Part of Problem Set \\ 4, 2\%\end{tabular} &
  - \\
 &
  \begin{tabular}[c]{@{}l@{}}e. Research   idea \\ presentation\end{tabular} &
  20\% &
  - \\ \hline
\multirow{3}{*}{Active} &
  \begin{tabular}[c]{@{}l@{}}f. Small group \\ discussion\end{tabular} &
  - &
  - \\
 &
  g. Debate &
  - &
  - \\
 &
  h. Peer advice &
  - &
  \begin{tabular}[c]{@{}c@{}}Final Presentations   \\ 6\%\end{tabular} \\ \hline
\multirow{2}{*}{Custom} &
  i. Python simulations &
  \begin{tabular}[c]{@{}c@{}}Problem Set 2 and \\ 3, each 10\%\end{tabular} &
  - \\
 &
  j. Game &
  - &
  \begin{tabular}[c]{@{}c@{}}Part of Problem Set \\ 4, 2\%\end{tabular} \\ \hline
  \multicolumn{4}{l}{Note: Grade weight is percent of final course grade determined by the assignment.}
\end{tabular}%
}
\end{table}

%% file: tables/Table2.tex
\begin{table}[H]
\centering
\caption{Engagement with Teaching Techniques}
\label{tab:2}
\resizebox{0.8\textwidth}{!}{%
\begin{tabular}{lL{8cm}}
\hline
\textbf{Item} &
  \textbf{Differences} \\ \hline
How   actively engaged &
  \begin{tabular}[c]{@{}L{8cm}@{}}\textbf{Lecture LESS} than Problem sets   (0.05), Quizzes / exams (0.03), Small group discussions (0.02)\\ \textbf{Written reflection   LESS}   than Lecture (0.04), Problem sets (\textless{}0.01), Quizzes / exams (\textless{}0.01),   Small group discussions (\textless{}0.01), Debate (0.02), Peer advice (0.01), Python   simulations (\textless{}0.01), Game (0.02)\end{tabular} \\ \hline 
  \multicolumn{2}{L{12.5cm}}{Note: \textit{p-values} in parentheses for pairwise Wilcoxon rank sum tests. Differences are reported in this table for all cases in which p $<$ 0.10. Students were asked “How actively did you engage in the following activities?” and gave responses for each activity on a 7-point scale, coded as -3 (Extremely inactive), -2 (Very inactive), -1 (Somewhat inactive), 0 (Neither), 1 (Somewhat active), 2 (Very active), 3 (Extremely active).}
\end{tabular}%
}
\end{table}

%% file: tables/Table3.tex
\begin{table}[H]
\centering
\caption{Enjoyment of Teaching Techniques}
\label{tab:3}
\resizebox{0.65\textwidth}{!}{%
\begin{tabular}{cL{9cm}}
\hline
\textbf{Item} &
  \textbf{Differences} \\ \hline
\multicolumn{1}{C{2cm}}{\vspace*{1.3cm} Enjoyableness} &
  \begin{tabular}[c]{@{}L{9cm}@{}}\textbf{Lecture LESS} than Game (0.02)\\
\textbf{Problem sets LESS} than Game ($<$ 0.01)\\
\textbf{Quizzes / exams LESS} than Lecture (0.04), Small group discussion (0.01), Game ($<$0.01) \\
\textbf{Written reflection LESS} than Game (0.01) \\ 
\textbf{Research idea presentation LESS} than Game (0.02) \\
\textbf{Debate LESS} than Game (0.03) \\
\textbf{Peer advice LESS} than Lecture ($<$0.01), Problem sets ($<$0.01), Written reflection ($<$0.01), Research idea presentation ($<$0.01), Small group discussion ($<$0.01), Debate ($<$0.01), Python simulation ($<$0.01)
Python simulation LESS than Game (0.02)
\end{tabular} \\ \hline 
  \multicolumn{2}{L{11.5cm}}{Note: \textit{p-values}  p-values in parentheses for pairwise Wilcoxon rank sum tests. Differences are reported in this table for all cases in which p < 0.10. Students ranked all ten techniques in response to the question “Which course activities did you enjoy the most?” Responses are rankings, with 10 being the highest rank and 1 being the lowest rank. }
\end{tabular}%
}
\end{table}

%% file: tables/Table4.tex
\begin{table}[H]
\centering
\caption{Reported Effectiveness of Teaching Techniques for Learning Objectives}
\label{tab:4}
\resizebox{0.7\textwidth}{!}{%
\begin{tabular}{lL{9.5cm}}
\hline
\textbf{\begin{tabular}[c]{@{}l@{}}Learning \\ objective\end{tabular}} &
  \textbf{Differences} \\ \hline
\begin{tabular}[c]{@{}l@{}}Remember \\ concepts\end{tabular} &
  \begin{tabular}[c]{@{}L{9.5cm}@{}}Written reflection   LESS   than Lecture (0.01), Problem sets (0.01), Quizzes / exams (0.02), Python   simulations (0.05)\\ Peer advice LESS than Lecture   (\textless{}0.01), Problem sets (\textless{}0.01), Quizzes / exams (\textless{}0.01), Research   idea presentation (0.02), Debate (0.01), Python simulations (0.01)\\ Game LESS than Lecture (0.02),   Problem sets (0.01), Quizzes / exams (0.02), Research idea presentation   (0.04), Debate (0.03)\end{tabular} \\
\begin{tabular}[c]{@{}l@{}}Understand different   \\ perspectives\end{tabular} &
  Peer advice LESS than Lecture (0.02),   Problem sets (0.02), Quizzes / exams (0.01), Written reflection (0.01),   Research idea presentation (0.02), Debate (\textless{}0.01), Python simulations   (0.01) \\
\begin{tabular}[c]{@{}l@{}}Analyze real-\\ world   issues\end{tabular} &
  \begin{tabular}[c]{@{}L{9.5cm}@{}}Lecture LESS than Research idea   presentation (0.05)\\ Problem sets LESS Written reflection   (0.04), Research idea presentation (0.01), Python simulations (0.02), Game   (0.03)   \\ Quizzes / exams LESS than Research idea   presentation (0.03)   \\ Peer advice LESS than Written   reflection (0.02), Research idea presentation (0.01), Small group discussion   (0.05), Debate (0.05), Python simulations (0.03), Game (0.01)\end{tabular} \\
\begin{tabular}[c]{@{}l@{}}Collaborative   \\ learning\end{tabular} &
  \begin{tabular}[c]{@{}L{9.5cm}@{}}Lecture LESS than Small group   discussion (0.02), Debate (0.03)\\ Problem sets LESS than Small group discussion (0.03)   \\ Quizzes / exams LESS than Small group   discussion (0.01), Debate (0.01) \\ Written reflection   LESS   than Small group discussion (\textless{}0.01), Debate (\textless{}0.01), Game (0.04)    \\ Research idea   presentation LESS   than Small group discussion (0.02), Debate (0.04)\\ Peer advice LESS than Small group   discussion (0.02)\\ Python simulation LESS than Small group discussion (0.01), Debate   (0.01)\\ Game LESS than Small group   discussion (0.02), Debate (0.01)\end{tabular} \\
\begin{tabular}[c]{@{}l@{}}Pros and cons of   \\ macroeconomic \\ policy\end{tabular} &
  \begin{tabular}[c]{@{}L{9.5cm}@{}}Problem sets LESS than Debate (0.05)\\ Written reflection   LESS   than Debate (\textless{}0.01)   \\ Research idea   presentation LESS   than Debate (\textless{}0.01)   \\ Small group   discussion LESS than   Debate (0.01)   \\ Peer advice LESS than Lecture (0.01),   Problem sets (\textless{}0.01), Quizzes / exams (0.02), Written reflection (0.01),   Research idea presentation (0.01), Debate (\textless{}0.01), Small group discussions   (0.03), Game (0.05)\\ Python simulation   LESS   than Lecture (0.01), Problem sets (0.02), Debate (\textless{}0.01)\end{tabular} \\ \hline
  \multicolumn{2}{L{12.5cm}}{Note: p-values in parentheses for pairwise Wilcoxon rank sum tests. Differences are reported in this table for cases in which p $<$ 0.10.  Data come from questions of the form “How effective do you think each of the following activities was at contributing to the learning objective of…” Responses are on a 7-point scale, coded as -3 (Extremely ineffective), -2 (Very ineffective), -1 (Somewhat ineffective), 0 (Neither), 1 (Somewhat effective), 2 (Very effective), 3 (Extremely effective).}
\end{tabular}%
}
\end{table}

%% file: tables/Table5.tex
\begin{table}[H]
\centering
\caption{Enjoyment of Teaching Techniques}
\label{tab:5}
\resizebox{0.8\textwidth}{!}{%
\begin{tabular}{cL{9cm}}
\hline
\textbf{Item} &
  \textbf{Differences} \\ \hline
\multicolumn{1}{C{2.2cm}}{\vspace*{1.2cm} Conduct systematic research} &
  \begin{tabular}[c]{@{}L{9cm}@{}} Written reflection LESS than Presentation (0.05), Lecture (0.05) \\
Game LESS than Presentation (0.01), Lecture (0.02), Python simulation (0.01)\\
Peer advice LESS than Written reflection (0.02), Presentation ($<$0.01), Debate (0.05), Lecture ($<$0.01), Python simulation ($<$0.01), Quizzes / exams (0.04), Problem sets (0.01), Small group discussion (0.03) \\
Quizzes / exams LESS than Presentation (0.03), Lecture (0.03), Python simulation (0.05)
\end{tabular} \\ \hline 
  \multicolumn{2}{L{11.5cm}}{Note: \textit{p-values}  in parentheses for pairwise Wilcoxon rank sum tests. Differences are reported in this table for cases in which p $<$ 0.10. Data come from the question “How effective do you think each of the following activities was at contributing to the learning objective of conducting macroeconomic research in a systematic way?” Responses were on a 7-point scale, coded as -3 (Extremely ineffective), -2 (Very ineffective), -1 (Somewhat ineffective), 0 (Neither), 1 (Somewhat effective), 2 (Very effective), 3 (Extremely effective).}
\end{tabular}%
}
\end{table}

%% file: tables/Table6.tex
\begin{table}[H]
\centering
\caption{Enjoyment of Teaching Techniques}
\label{tab:6}
\resizebox{0.8\textwidth}{!}{%
\begin{tabular}{cL{9cm}}
\hline
\textbf{Item} &
  \textbf{Differences} \\ \hline
\multicolumn{1}{C{2.5cm}}{\vspace*{1.8cm} Usefulness in developing research} &
  \begin{tabular}[c]{@{}L{9cm}@{}} Lecture LESS than Research idea presentation (0.03), Python simulation (0.01) \\
Problem sets LESS than Python simulation (0.03)\\
Quizzes / exams LESS than Lecture ($<$0.01), Problem sets ($<$0.01), Written reflection ($<$0.01), Research idea presentation ($<$0.01), Small group discussion ($<$0.01), Debate ($<$0.01), Python simulation ($<$0.01), Game ($<$0.01)\\
Small group discussion LESS than Python simulation (0.03)\\
Debate LESS than Python simulation (0.04)\\
Peer advice LESS than Lecture (0.01), Problem sets ($<$0.01), Written reflection ($<$0.01), Research idea presentation ($<$0.01), Small group discussion ($<$0.01), Debate (0.02), Python simulation ($<$0.01), Game (0.01)\\
Game LESS than Python simulation (0.04)
\end{tabular} \\ \hline 
  \multicolumn{2}{L{12cm}}{Note: \textit{p-values}  in parentheses for pairwise Wilcoxon rank sum tests. Differences are reported in this table for all cases in which p $<$ 0.10. Students ranked each of the 10 activities in order of how useful they were in helping them develop their Signature Work, which is the name in this academic program for a student’s culminating project, required for the degree. Responses are rankings, with 10 being the highest rank and 1 being the lowest rank.}
\end{tabular}%
}
\end{table}

%% file: tables/Table7.tex
\begin{table}[H]
\centering
\caption{Correlation between Engagement and Learning Objectives; Panel 1}
\label{tab:7}
\resizebox{0.8\textwidth}{!}{%
\begin{tabular}{cccc}
\hline
\textbf{\begin{tabular}[c]{@{}c@{}}Learning \\ Objectives\end{tabular}} &
  \textbf{Learning Techniques} &
  \textbf{\begin{tabular}[c]{@{}c@{}}Spearman Correlation\\ CI95\%\end{tabular}} &
  \textbf{\begin{tabular}[c]{@{}c@{}}Spearman Correlation\\ p-value\end{tabular}} \\ \hline
\multirow{10}{*}{\begin{tabular}[c]{@{}c@{}}Remember   \\ concepts\end{tabular}}       & Lecture                    & {[}0.33, 0.81{]} & 0.00 \\
                                                                                       & Problem sets               & {[}0.41, 0.84{]} & 0.00 \\
                                                                                       & Quizzes / exams            & {[}0.34, 0.81{]} & 0.00 \\
                                                                                       & Written reflection         & {[}0.37, 0.82{]} & 0.00 \\
                                                                                       & Research idea presentation & {[}0.32, 0.8{]}  & 0.00 \\
                                                                                       & Small group discussion     & {[}0.19, 0.75{]} & 0.00 \\
                                                                                       & Debate                     & {[}0.47, 0.86{]} & 0.00 \\
                                                                                       & Peer advice                & {[}0.25, 0.78{]} & 0.00 \\
                                                                                       & Python simulation          & {[}0.53, 0.88{]} & 0.00 \\
                                                                                       & Game                       & {[}0.6, 0.9{]}   & 0.00 \\ \hline
\multirow{10}{*}{\begin{tabular}[c]{@{}c@{}}Understand  \\  perspectives\end{tabular}} & Lecture                    & {[}0.35, 0.82{]} & 0.00 \\
                                                                                       & Problem sets               & {[}0.46, 0.86{]} & 0.00 \\
                                                                                       & Quizzes / exams            & {[}0.51, 0.87{]} & 0.00 \\
                                                                                       & Written reflection         & {[}0.38, 0.83{]} & 0.00 \\
                                                                                       & Research idea presentation & {[}0.1, 0.71{]}  & 0.02 \\
                                                                                       & Small group discussion     & {[}0.43, 0.85{]} & 0.00 \\
                                                                                       & Debate                     & {[}0.3, 0.8{]}   & 0.00 \\
                                                                                       & Peer advice                & {[}0.24, 0.78{]} & 0.00 \\
                                                                                       & Python simulation          & {[}0.55, 0.89{]} & 0.00 \\
                                                                                       & Game                       & {[}0.4, 0.84{]}  & 0.00 \\ \hline
\end{tabular}%
}
\end{table}

%% file: tables/Table8.tex
\begin{table}[H]
\centering
\caption{Correlation between Engagement and Learning Objectives; Panel 2}
\label{tab:8}
\resizebox{0.8\textwidth}{!}{%
\begin{tabular}{cccc}
\hline
\textbf{\begin{tabular}[c]{@{}c@{}}Learning \\ Objectives\end{tabular}} &
  \textbf{Learning Techniques} &
  \textbf{\begin{tabular}[c]{@{}c@{}}Spearman Correlation  \\ CI95\%\end{tabular}} &
  \textbf{\begin{tabular}[c]{@{}c@{}}Spearman Correlation\\ p-value\end{tabular}} \\ \hline
\multirow{10}{*}{\begin{tabular}[c]{@{}c@{}}Analyze real-\\ world issues\end{tabular}} & Lecture                    & {[}0.31, 0.8{]}  & 0.00 \\
                                                                                       & Problem sets               & {[}0.28, 0.79{]} & 0.00 \\
                                                                                       & Quizzes / exams            & {[}0.32, 0.81{]} & 0.00 \\
                                                                                       & Written reflection         & {[}0.26, 0.78{]} & 0.00 \\
                                                                                       & Research idea presentation & {[}0.1, 0.71{]}  & 0.01 \\
                                                                                       & Small group discussion     & {[}0.28, 0.79{]} & 0.00 \\
                                                                                       & Debate                     & {[}0.27, 0.79{]} & 0.00 \\
                                                                                       & Peer advice                & {[}0.2, 0.76{]}  & 0.00 \\
                                                                                       & Python simulation          & {[}0.32, 0.8{]}  & 0.00 \\
                                                                                       & Game                       & {[}0.63, 0.91{]} & 0.00 \\ \hline
\multirow{10}{*}{\begin{tabular}[c]{@{}c@{}}Collaborative\\  learning\end{tabular}}    & Lecture                    & {[}0.23, 0.77{]} & 0.00 \\
                                                                                       & Problem sets               & {[}0.32, 0.8{]}  & 0.00 \\
                                                                                       & Quizzes / exams            & {[}0.32, 0.8{]}  & 0.00 \\
                                                                                       & Written reflection         & {[}0.31, 0.8{]}  & 0.00 \\
                                                                                       & Research idea presentation & {[}0.15, 0.73{]} & 0.01 \\
                                                                                       & Small group discussion     & {[}0.4, 0.84{]}  & 0.00 \\
                                                                                       & Debate                     & {[}0.22, 0.77{]} & 0.00 \\
                                                                                       & Peer advice                & {[}0.41, 0.84{]} & 0.00 \\
                                                                                       & Python simulation          & {[}0.31, 0.8{]}  & 0.00 \\
                                                                                       & Game                       & {[}0.49, 0.87{]} & 0.00 \\ \hline
\end{tabular}%
}
\end{table}

%% file: tables/Table9.tex
\begin{table}[H]
\centering
\caption{Correlation between Engagement and Learning Objectives; Panel 3}
\label{tab:9}
\resizebox{0.8\textwidth}{!}{%
\begin{tabular}{cccc}
\hline
\textbf{\begin{tabular}[c]{@{}c@{}}Learning \\ Objectives\end{tabular}} &
  \textbf{Learning Techniques} &
  \textbf{\begin{tabular}[c]{@{}c@{}}Spearman Correlation\\ CI95\%\end{tabular}} &
  \textbf{\begin{tabular}[c]{@{}c@{}}Spearman Correlation\\ p-value\end{tabular}} \\ \hline
\multirow{10}{*}{\begin{tabular}[c]{@{}c@{}}Pros \& cons of \\ monetary policy\end{tabular}} &
  Lecture &
  {[}0.25, 0.78{]} &
  0.00 \\
                                     & Problem sets               & {[}0.52, 0.88{]}  & 0.00 \\
                                     & Quizzes / exams            & {[}0.28, 0.79{]}  & 0.00 \\
                                     & Written reflection         & {[}0.28, 0.79{]}  & 0.00 \\
                                     & Research idea presentation & {[}-0.04, 0.63{]} & 0.07 \\
                                     & Small group discussion     & {[}0.22, 0.76{]}  & 0.00 \\
                                     & Debate                     & {[}0.41, 0.84{]}  & 0.00 \\
                                     & Peer advice                & {[}0.27, 0.79{]}  & 0.00 \\
                                     & Python simulation          & {[}0.44, 0.85{]}  & 0.00 \\
                                     & Game                       & {[}0.58, 0.89{]}  & 0.00 \\ \hline
\multirow{10}{*}{Conduct   research} & Lecture                    & {[}0.25, 0.78{]}  & 0.00 \\
                                     & Problem sets               & {[}0.43, 0.85{]}  & 0.00 \\
                                     & Quizzes / exams            & {[}0.27, 0.79{]}  & 0.00 \\
                                     & Written reflection         & {[}0.26, 0.78{]}  & 0.00 \\
                                     & Research idea presentation & {[}0.29, 0.79{]}  & 0.00 \\
                                     & Small group discussion     & {[}0.2, 0.76{]}   & 0.00 \\
                                     & Debate                     & {[}0.39, 0.83{]}  & 0.00 \\
                                     & Peer advice                & {[}0.21, 0.76{]}  & 0.00 \\
                                     & Python simulation          & {[}0.48, 0.86{]}  & 0.00 \\
                                     & Game                       & {[}0.55, 0.88{]}  & 0.00 \\ \hline
\multicolumn{4}{L{15cm}}{Note: CI95\% and p-value are from Spearman correlation between student’s answer to how actively they engaged with a technique and how effective they reported it was with regard to the named learning objective. Techniques are only listed for an outcome if the correlation is statistically significant.}
\end{tabular}%
}
\end{table}

%% file: tables/Table10.tex
\begin{table}[H]
\centering
\caption{Correlation between Enjoyment and Learning Objectives}
\label{tab:10}
\resizebox{0.8\textwidth}{!}{%
\begin{tabular}{cccc}
\hline
\textbf{\begin{tabular}[c]{@{}c@{}}Learning \\ Objectives\end{tabular}} &
  \textbf{Learning Techniques} &
  \textbf{\begin{tabular}[c]{@{}c@{}}Spearman Correlation\\ CI95\%\end{tabular}} &
  \textbf{\begin{tabular}[c]{@{}c@{}}Spearman Correlation\\ p-value\end{tabular}} \\ \hline
\begin{tabular}[c]{@{}c@{}}Remember\\  concepts\end{tabular}                                & Peer advice     & {[}0.19, 0.75{]}   & 0.00 \\ \hline
\begin{tabular}[c]{@{}c@{}}Understand \\ perspectives\end{tabular}                          & Peer advice     & {[}0.19, 0.75{]}   & 0.00 \\ \hline
\begin{tabular}[c]{@{}c@{}}Analyze real-\\ world issues\end{tabular}                        & Peer advice     & {[}0.04, 0.68{]}   & 0.03 \\ \hline
\begin{tabular}[c]{@{}c@{}}Collaborative \\ learning\end{tabular}                           & Quizzes / exams & {[}-0.73, -0.14{]} & 0.01 \\ \hline
\multirow{3}{*}{\begin{tabular}[c]{@{}c@{}}Pros \& cons of\\  monetary policy\end{tabular}} & Quizzes / exams & {[}-0.66, -0.01{]} & 0.04 \\
                                                                                            & Peer advice     & {[}0.14, 0.73{]}   & 0.01 \\
                                                                                            & Game            & {[}0.06, 0.69{]}   & 0.03 \\ \hline
\multirow{2}{*}{Conduct research}                                                           & Quizzes / exams & {[}-0.67, -0.03{]} & 0.04 \\
                                                                                            & Peer advice     & {[}0.11, 0.72{]}   & 0.01 \\ \hline
\multicolumn{4}{L{15cm}}{Note: CI95\% and p-value are from Spearman correlation between student’s answer to how actively they engaged with a technique and how effective they reported it was with regard to the named learning objective. Techniques are only listed for an outcome if the correlation is statistically significant.}
\end{tabular}%
}
\end{table}

%% file: tables/Table11.tex
\begin{table}[H]
\centering
\caption{Correlation between Engagement and Enjoyment}
\label{tab:11}
\resizebox{0.8\textwidth}{!}{%
\begin{tabular}{ccc}
\hline
\textbf{Learning Techniques} &
  \textbf{\begin{tabular}[c]{@{}c@{}}Spearman Correlation\\ CI95\%\end{tabular}} &
  \textbf{\begin{tabular}[c]{@{}c@{}}Spearman Correlation\\ p-value\end{tabular}} \\ \hline
Research   idea presentation &
  {[}0.02, 0.66{]} &
  0.04 \\ \hline 
  \multicolumn{3}{L{12cm}}{Note: CI95\% and p-value are from Spearman correlation between student’s answer to how actively they engaged with a technique and how much they reported enjoying it. Techniques are only listed for an outcome if the correlation is statistically significant.}
\end{tabular}%
}
\end{table}